\documentstyle[aps,12pt,manuscript]{revtex}

\begin{document}

\def\be{\begin{equation}}
\def\ee{\end{equation}}
\def\bea{\begin{eqnarray}}
\def\eea{\end{eqnarray}}

\def\pd{\partial}
\def\a{\alpha}
\def\b{\beta}
\def\g{\gamma}
\def\d{\delta}
\def\m{\mu}
\def\n{\nu}
\def\t{\tau}
\def\l{\lambda}

\def\s{\sigma}
\def\e{\epsilon}
\def\scri{\mathcal{J}}
\def\cM{\mathcal{M}}
\def\tcM{\tilde{\mathcal{M}}}
\def\RR{\mathbb{R}}

\preprint{INJE-TP-99-1}

\title{6D black string as a model of the AdS/CFT correspondence}

\author{Y. S. Myung, N.J. Kim and H.W. Lee}
\address{Department of Physics, Inje University, Kimhae 621-749, Korea} 

\maketitle

\begin{abstract}
We discuss the entropy for the extremal BTZ black hole and 
the extremal EBTZ black hole. 
The EBTZ black hole means the BTZ black hole 
embedded in a five-dimensional(5D) black hole.  
The six-dimensional(6D) black string with traveling waves is introduced 
as a concrete model for realizing the AdS/CFT correspondence. 
The traveling waves carry the momentum distribution $p(u)$ which 
plays an important role in counting the entropy and 
establishing the correspondence.
It turns out that the EBTZ black hole 
is consistent with the AdS/CFT correspondence. 
\end{abstract}

\newpage

\section{Introduction}
Recently the AdS/CFT correspondence has attracted much interest. 
This is based on the duality relation between the string theory
(bulk theory) on AdS$_{d+1}$ and a conformal field theory(CFT) as its 
d-dimensional boundary theory\cite{Mal9711200,Gub9802109,Wit9802150}. 
The s-wave calculation of greybody factor (dynamic property) for the 
BTZ black hole on AdS$_3$ is in agreement with that of a 5D black hole 
with three charges $Q_1, Q_5 , N_K $ \cite{Lee98PRD104013}. This is 
obvious because in the low energy limit a 5D black hole reduces 
to the BTZ black hole\cite{Ske9901050}. 
Furthermore the greybody factor for the BTZ black hole agrees with the 
CFT$_2$ calculation on the boundary of AdS$_3$\cite{Bir97PLB281}. This 
proves the AdS$_3$/CFT$_2$ correspondence for the greybody factor. 
In the calculation of the greybody factor we need to study the propagation 
of test fields in the BTZ black hole background\cite{Lee98PRD084022}. 
For counting the entropy we investigate the static configuration 
of the black hole itself\cite{Str96PLB99,Hor9605224}. 
A 5D black hole was originally 
constructed by the bound states of $Q_1$ D1-branes and $Q_5$ D5-branes 
with some momentum $P(=2 \pi N_K / L = N_K / R_1)$ along 
an internal circle(S$^1$) 
with the radius $R_1$. In order to see a 
closer relation between the microscopic states and the classical spacetime, 
it is convenient to rewrite a 5D black hole 
to include S$^1$ as a space. It comes out a six-dimensional(6D) black string. 
Even though the extension of the spacetime, 
the entropy for a 5D black hole is the same as in a 6D black string. 
In this work a 6D black string with traveling waves is introduced 
as a concrete model for realizing the AdS/CFT correspondence. 
The traveling waves carry the momentum distribution. This 
plays an important role in counting the entropy and 
establishing the AdS/CFT correspondence.

Nowdays it seems to be a discrepancy for counting the entropy 
(static property) of the BTZ black hole in relation to the AdS/CFT 
correspondence. First we assume that the BTZ black hole is not 
embedded in string theory. 
In this sense this corresponds to an exact AdS$_3$ background. 
And thus this is described as a solution 
to the pure three-dimensional(3D) gravity($S_{3D} = {1 \over {16 \pi G_3}} \int 
d^3 x \sqrt{-g} ( R- 2 \Lambda)$) with the negative cosmological 
constant ($\Lambda = - 1/ \ell^2$)
\begin{equation}
ds_{\rm BTZ}^2 = - {{(\rho^2 - \rho_+^2) ( \rho^2 - \rho_-^2) } \over 
  {\rho^2 \ell^2}} dt^2 + \rho^2 \left ( d \phi - { J \over {2 \rho^2}} dt 
\right )^2 + {{\rho^2 \ell^2} \over 
{(\rho^2 -\rho_+^2) ( \rho^2 - \rho_-^2)}} d\rho^2 .
\label{metric-BTZ}
\end{equation}
Here $M=(\rho_+^2 + \rho_-^2) / 8 G_3 \ell^2$ and 
$J = \rho_+ \rho_-/ 4 G_3 \ell$ correspond to the mass and angular momentum 
of the BTZ black hole. $\rho_+(\rho_-)$ are the outer(inner) horizon.
This can be described by a SL(2,{\bf R})$_L \times $SL(2,{\bf R})$_R$ 
Chern-Simons theory. Since it is a topological field theory, 
the physical degrees of freedom reside only on the boundary. 
In this direction an important question arises: what fields provide 
the relevant degrees of freedom on the boundary at infinity($\rho=\infty$)? 
A single Liouville field at infinity may be a candidate.  
Unfortunately this has a 
central charge($c=1$\cite{Kut91NPB600,Mar9809021}). But we need 
$c=3 \ell / 2 G_3$\cite{Bro86CMP207,Sei90PTPSB319} 
for AdS$_3$/CFT$_2$ correspondence at $\rho = \infty$. 
That is, in order to recover the Bekenstein-Hawking entropy of the 
extremal BTZ black hole with $\rho_-=\rho_+=\rho_0$
\begin{equation}
{\rm S}^{\rm BTZ}_{\rm BH} = { 2 \pi \rho_0 \over 4 G_3},
\label{entropy-BH}
\end{equation}
one needs both $c = 3 \ell / 2 G_3$ and Cardy's formula
\begin{equation}
d(c,N) = \exp \left ( 2 \pi \sqrt{{c \over 6} N } \right )
\label{cardy}
\end{equation}
with the level number $N \gg 1$.
Here $N= \rho_0^2 / 4 G_3 \ell$ is given by an eigenvalue of the 
angular momentum operator $J=L_0$. At this stage we ask : 
is this BTZ black hole dual to the CFT of the D1-D5 bound states
(D-string theory)? Because it is not embedded in string theory, its 
boundary CFT with $c=1$ has nothing to do with the D-string. 
The Liouville theory is simply a boundary description of the pure 
3D gravity sector. 
Martinec showed that the 3D gravity is a pure 
gauge theory and thus it examines only its macroscopic properties 
(thermodynamics) by a set of Noether charges \cite{Mar9809021}. 
On the other hand, the gauge theory of branes(dual CFT$_2$) 
is a tool to investigate 
its microscopic features. Hence the 3D gravity 
appears as a collective field theory of the microscopic dual CFT$_2$.
On the other hand, string theory provides many more dynamical fields
($c= 6 Q_1 Q_5$) and one certainly does not expect the Liouville 
sector to accommodate all of the states of the system\cite{Ske9901050}. 
Hence it is not appropriate for discussing the AdS/CFT correspondence 
for the entropy within the pure 3D gravity.
In this paper, we wish to describe how the AdS/CFT correspondence for 
the entropy can be realized explicitly. 

\section{The EBTZ black hole}
Now we consider the BTZ black hole 
which resides in the near-horizon geometry of a 5D black hole. Hereafter 
we call this the embedded BTZ(EBTZ) 
black hole\cite{Mal9711200,Lee98PRD104013,Mal9804085,Sat9810135}. 
At low energies ($\alpha' \to 0$) the physics of an extremal 5D black hole 
(6D black string) is governed by the extremal EBTZ black hole. 
In this case we have an embedded AdS$_3$. Explicitly a 
5D black hole(M$_5$) has the geometry : AdS$_3$ in the near-horizon 
(the throat region) but with asymptotically flat space. 
This corresponds to the geometry which interpolates between 
AdS$_3$ and flat space. This black hole has 
the outer horizon at $r=r_0$ and the inner horizon at $r=0$. And thus 
the extremal limit means $r=r_0 \to r=0$. In this limit the 
coordinate relation 
between a 5D black hole and the EBTZ black hole is given by 
$\tilde \rho^2 = r^2 + \tilde \rho_0 ^2 $, 
$\tilde \phi = z/R_1, \tilde t = t \tilde Q_1 \tilde Q_5/R_1^2$
The extremal EBTZ black hole 
has its horizon at $\tilde \rho = \tilde \rho_0$. If one starts with 
M$_5 \times$S$^1 \times$T$^4$ in the ten-dimensional(10D) 
type IIB superstring theory, 
we have AdS$_3 \times $S$^3 \times $T$^4$ in the throat region 
($r_0 \le r \le R, R^2 = \sqrt{\tilde Q_1 \tilde Q_5}$) 
but find Minkowski space after 
passing through the remote boundary ($r=R$). Hereafter we choose a 
convention with $\alpha'=1$. In the extremal limit the relations 
between $\tilde Q_i$ and $Q_1, Q_5, N_K$ are given by 
$v \tilde Q_1 = g_s Q_1$, $\tilde Q_5 = g_s Q_5$, and 
$R_1^2 v \tilde Q_K = g_s^2 N_K$.  The 3D Newton's 
constant($G_3$) is replaced by \cite{Ske9901050}
\begin{equation}
\tilde G_3 = {g_s^2 \over 4 v R_1 R^2 }
\label{G-3D}
\end{equation}
where $g_s$ is the string coupling constant and $v = V/(2 \pi)^4$ with the 
volume of T$^4$($V$). 
Notice that $\tilde G_3$ contains all information of the type IIB 
string theory, contrary to $G_3$. 
The geometry of the extremal EBTZ black hole is described by
\begin{equation}
d \tilde s_{\rm EBTZ}^2 = 
-{(\tilde \rho^2 -\tilde \rho_0^2)^2 \over \tilde \rho^2 \tilde \ell^2 } 
d \tilde t^2 
+\tilde \rho^2 \left ( d \tilde \phi - {\tilde J \over 2 \tilde \rho^2}
d \tilde t \right ) ^2 
+ { {\tilde \rho^2 \tilde \ell^2} \over (\tilde \rho^2 -\tilde \rho_0^2 ) ^2 }
d \tilde \rho^2.
\label{metric-EBTZ}
\end{equation}
The mass, angular momentum, and the area of the horizon take the forms
\begin{equation}
\tilde M = \tilde J / \tilde \ell,~~ 
\tilde J = { \tilde \rho_0 \over 4 \tilde G_3 \tilde \ell }, ~~
\tilde A_3 = 2 \pi \tilde \rho_0 = 2 \pi \sqrt{\tilde Q_K}
\label{mass-EBTZ}
\end{equation}
with $ \tilde \ell = R^2 / R_1, \tilde \rho_0^2 = \tilde Q_K =
g_s^2 N_K / R_1^2 v$ .
Hence the Bekenstein-Hawking entropy of the EBTZ black hole leads to 
\begin{equation}
{\rm S}^{\rm EBTZ}_{\rm BH} = { \tilde A_3 \over 4 \tilde G_3 } 
= 2 \pi \sqrt{Q_1 Q_5 N_K } = { A_5 \over 4 G_5} 
= {\rm S}^{\rm 5D}_{\rm BH}
\label{entropy-EBTZ}
\end{equation}
which is exactly the same form as in the 5D black hole. Therefore all 
information is encoded in AdS$_3$.

Suppose the weakly coupled D-brane description in which 
BPS states are described by right movers with $N_K$ 
momentum in (1+1) dimensions. Then 
the degeneracy of D-brane system is given by the degeneracy of CFT with 
a central charge $\tilde c = 3 \tilde \ell / 2 \tilde G_3 = 6 Q_1 Q_5$ at 
level $N_K$. Plugging this into (\ref{cardy}) leads to the 
Bekenstein-Hawking entropy (\ref{entropy-EBTZ}). However, this is done actually 
in the weak-coupling limit of $g_sQ_1, g_sQ_5, g_s^2 N_K \ll 1$. This is 
precisely the opposite regime where the classical supergravity solution 
is good. On the supergravity side a 5D black hole appears in the strong 
coupling limit with $g_sQ_1, g_sQ_5, g_s^2 N_K \gg 1$. Due to 
supersymmetry, one can extrapolate results in the D-brane phase to the 
black hole phase.

\section{6D black string}
\subsection{6D black string without momentum modes}
We start with a 6D black string with $\tilde Q_1 = \tilde Q_5, p=0$ as
\cite{Tay9806132}
\begin{equation}
ds^2_{\rm 6D} = - \left ( 1 + {\tilde Q_1 \over r^2 } \right )^{-1} 
 du dv  + \left ( 1 + {\tilde Q_1 \over r^2 } \right )
\left (  dr^2 + r^2 d\Omega_3^2 \right ),
\label{metric-6DN0}
\end{equation}
where $u=t-z$ and $v=t+z$. $z$ is on S$^1$.
It turns out to be AdS$_3\times$S$^3$
in the near-horizon
\begin{eqnarray}
d{\tilde s}_{\rm 6D}^2 &=& 
- {r^2 \over R^2} du dv + {R^2 \over r^2} dr^2  + R^2 d \Omega_3^2 
\nonumber \\
&=&{R^2 \over y^2} \left [-du dv + dy^2 \right ] + R^2 
d \Omega_3^2 , ~~  y=R^2/r.
\label{metric-6DNH}
\end{eqnarray} 
Here $R$ is the effective radius of an embedded AdS$_3$.
This amounts to the zero temperature limit of the D-string 
model with $Q_1$ and $Q_5$. Although this is a simple model for 
the calculation of the greybody factor, it is not good for the 
entropy counting. The EBTZ black hole is a rotating black hole with the 
angular momentum $\tilde J$. This contains the information for the 
momentum modes. The important thing is how we introduce the total 
momentum number $N(=N_K)$ into (\ref{metric-6DN0}).

\subsection{6D black string with $p = {\rm\bf constant}$}
Recently it is shown that the chiral orbifold of AdS$_3$/Z$_N$ describes 
the extremal EBTZ black hole 
of (\ref{metric-EBTZ})\cite{Mal9804085,Beh9809015}. 
Here the chiral orbifold of 
AdS$_3$/Z$_N$ means a lens space of AdS$_3$\cite{Duf9807173}. 
Also (AdS$_3$/Z$_N$) $\times$ S$^3$ corresponds to the near-horizon 
geometry of the extremal 6D black string with momentum modes 
$p =\tilde \rho_0^2 = g_s^2 N_K / R_1^2 v \equiv c N, 
 c = 2 \pi \kappa^2 / L^2$\cite{Lee98PRD104013},
\begin{equation}
d{\tilde s}_{{\rm AdS}_3/{\rm Z}_N}^2 = {r^2 \over R^2} \left [ -du dv 
   + { cN \over r^2} du^2 \right ] + {R^2 \over r^2} dr^2 
   + R^2 d \Omega_3^2.
\label{metric-AdSP}
\end{equation}
Here $\kappa^2 = 2 \pi g_s^2 / v$ is related to the 6D Newton's constant. 
This corresponds to a homogeneous (translationally invariant) black string.
(\ref{metric-AdSP}) contains the geometry of the EBTZ black hole 
in (\ref{metric-EBTZ}). The difference is that $N$-term is explicitly 
shown here in terms of $r$-coordinate. 
However, the horizon geometry ($\bar g^{rr} = 0$) is not changed 
under the coordinate transformation $r \leftrightarrow \tilde \rho$. 
Consequently, its entropy 
counting goes exactly to the same way as in the EBTZ black hole.
This guarantees that the entropy of the EBTZ black hole is 
invariant under this orbifolding of
${\rm AdS}_3 \to {\rm AdS}_3/{\rm Z}_N$.
At this stage let us introduce the AdS/CFT correspondence. This means that the 
10D type IIB bulk theory deep in the throat is dual to 
the 2D CFT of D-string(world sheet theory) on its remote boundary. 
Further, this implies that the semiclassical limit of spacetime 
physics (a homogeneous black string) is related to the large 
($Q_1, Q_5, N$) limit of the CFT(D-string).
Concerning the entropy, the relevant quantity is the 
total momentum number $N$. Furthermore $Q_1$ and $Q_5$ are 
fixed numbers. Considering $N$ as a variable, one can 
understand clearly how this correspondence is realized.

\subsection{6D black string with traveling waves $p(u)$}
We wish to describe how the AdS/CFT correspondence can 
be realized in this picture. 
First we extend a black string to include traveling waves. 
The extremal 6D black string solution (\ref{metric-6DN0}) 
has full (1+1)-dimensional Poincar\'e invariance, including a null 
translational symmetry. Thanks to this null symmetry, one can add traveling 
waves with the momentum distribution $p(u)$ to any solution as
\begin{equation}
{ds^p_{\rm 6D}}^2 = \left ( 1 + {R^2 \over r^2 } \right )^{-1} 
\left ( - du dv + {p(u) \over r^2} du^2 \right ) + 
\left ( 1 + {R^2 \over r^2 } \right )
\left (  dr^2 + r^2 d\Omega_3^2 \right ).
\label{metric-6D}
\end{equation}
Although $p(u)$ looks like a pure gauge degree of freedom, it contains 
an important physical information (momentum distribution). 
Furthermore $p(u)$ carries this information far from the horizon. 
The mechanism is as follows. In the low energy limit 
the 6D black string without $p(u)$ becomes AdS$_3\times$S$^3$. 
That is, (\ref{metric-6DN0}) leads to (\ref{metric-6DNH}). 
Hence a naive counting of the graviton ($D(D-3)/2$)
implies that in (2+1)-dimensional spacetime we have no 
propagating graviton. 
This counting is suitable for the black holes 
and non-extremal black strings. However, our model is the extremal black 
string with the null Killing symmetry. In this case the graviton 
mode ($p(u)$) may be physically propagating. Actually this graviton 
becomes a propagating mode by the transmutation of the degree of 
freedom with the other field (for example, dilaton) in the extremal 
black string spacetime\cite{Lee97MPLA545}. 
This is similar to the Higgs mechanism for gauge 
fields in the Minkowski spacetime. Therefore this graviton 
is not traveling along the horizon but becomes purely outgoing 
in the near-horizon.  This is longitudinal wave moving along the string. 
It is important to note that 
this wave propagates indefinitely without radiating to infinity or falling 
into the horizon. 

Further we review briefly how these traveling waves can be 
generated. This is found from the technique 
of Garfinkle-Vachaspati\cite{Gar90PRD1960}.  Starting from a known 
static solution, this produces a new solution 
representing waves traveling on the old black string background. 
However, it requires that the 
background metric ($\bar g_{MN}$) possess a null, orthogonal 
Killing vector($k^M = ( \partial/ \partial v)^M$). As an example, 
if ($\bar g_{MN} , \bar H_{MNP}$) is also an exact solution to the 6D 
Einstein-Maxwell equations, 
then ($\bar g_{MN} + h_{MN} , \bar H_{MNP}$) is also an exact solution to 
this system\cite{Lee97MPLA545}. 
Here $h_{MN} = hk_M k_N$ with $ h\bar g_{vu} = p(u)/r^2$. 
This method was 
originally proposed for the Yang-Mills-Higgs system coupled to gravity 
and then applied to the low energy limit of string theory
\cite{Car98PRD104019}

The Bekenstein-Hawking entropy for a 5D black 
hole and the EBTZ black hole can also be reproduced by this 
picture\cite{Hor9605224}, provided only that the distribution function $p(u)$ 
does not vary too rapidly : $p^{3/2} \gg r_0^2 \vert \dot p \vert$, where 
the dot means $d/du$. The geometry of the near-horizon for
the 10D spacetime is
\begin{equation}
d \tilde s_{\rm 10D}^2 = { r^2 \over R^2} \left [ - du dv +
{p(u) \over r^2} du^2 + { R^4 \over r^4} dr^2 \right ] 
+ R^2 d \Omega_3^2 + d y_i^2
\label{metric-10D}
\end{equation}
with $y^i$ on the T$^4$. 
We note that this is neither AdS$_3 \times$S$^3 \times$T$^4$ nor 
(AdS$_3$/Z$_N$)$\times$S$^3 \times$T$^4$.
In order to rewrite this as 
AdS$_3\times$S$^3 \times $T$^4$, we must introduce a function $\sigma$, 
periodic in $u$ which is related to $p(u)$ by 
$\sigma^2 + \dot \sigma = p/r_0^4$. 
Then the near-horizon geometry can be 
described by AdS$_3 \times$S$^3 \times$T$^4$
and the horizon area  is given by 
$A_{10} = 2 \pi^2 r_0^2 V \int_0^L \sigma(u) du$.
If $p^{3/2} \gg r_0^2 |\dot p |$, one has $\sigma = \sqrt{p/r_0^4}$.
Its Bekenstein-Hawking entropy is
\begin{equation}
{\rm S}^{\rm 10D}_{\rm BH} = {A_{10} \over 4 G_{10} } =
\sqrt{2 \pi Q_1 Q_5 } \int_0^L \sqrt{ p(u) \over \kappa^2} du .
\label{entropy-10}
\end{equation}
If one fixes the total momentum, the horizon area is 
maximized by the uniform distribution of $p={\rm constant}$. Then one finds 
a homogeneous (translationally invariant) black string. In this case 
we have $p/ \kappa^2 = P/L = 2 \pi N /L^2 $ and thus 
(\ref{entropy-10}) reduces to the familiar form of 
S$^{\rm 5D}_{\rm BH}= 2 \pi \sqrt{Q_1 Q_5 N} = $S$^{\rm EBTZ}_{\rm BH}$. 
On the other hand, fixing the momentum distribution 
$p(u)$, one finds the form of (\ref{entropy-10}).

We wish to count BPS states. 
All of string fields should have the same 
momentum distribution as in the black string: 
$p^{3/2} \gg r_0^2 \vert \dot p \vert$. 
Quantum mechanically, however, one cannot fix the momentum distribution 
$p(u)$ of string fields exactly in  D-string theory. 
This is so because $p(u)$ and $p(u')$ do not commute and thus 
its Fourier modes satisfy the Virasoro algebra. 
Instead one introduces a mesoscopic length scale 
$\ell_m \ll L$. One divides the spacetime into 
$X=L/\ell_m$ intervals $\Delta_a(a \in \{ 1, \cdots, X \} )$.
Then the total momentum in the interval $\Delta_a$ is 
$P_a = \kappa^{-2} \int_{\Delta_a} p(u) du$ and 
it is assumed to be constant.
If one could view states on S$^1$ as consisting of a collection 
of $X$-independent systems of length $\ell_m$, the number of 
states for $\tilde c = 6Q_1 Q_5$ string fields with total momentum 
$P_a = 2 \pi N_a / \ell_m$ on the $a$-th interval would be 
$e^{S_a}$ with $S_a= 2 \pi \sqrt{\tilde c N_a / 6 } =
\sqrt{2 \pi Q_1 Q_5 P_a \ell_m}$. Here $N_a \gg Q_1 Q_5$, because of 
$p/|\dot p| \gg \ell_m \gg \sqrt{Q_1 Q_5 \kappa^2/p}$. 
The entropy of $X$-independent systems is additive so the total entropy 
with $P_1, \cdots, P_X$ is given by
\begin{equation}
{\rm S}^p_{\rm CFT} = 2 \pi \sqrt{Q_1 Q_5} \sum_{a=1}^X \sqrt{N_a} = 
\sqrt{2 \pi Q_1 Q_5} \int_0^L \sqrt{ { p \over \kappa^2}} du
\label{entropy-P}
\end{equation}
which leads to (\ref{entropy-10}).

We note again that on the black string side the traveling waves play 
an important role in transferring the information of the horizon 
into the remote boundary at $r=R$. 
The AdS/CFT correspondence means that the semiclassical 
limit of spacetime physics (a 6D black string with $p(u)$) is 
related to the large ($Q_1, Q_5, p(u)/\kappa^2$) limit of the CFT(D-string).
We propose here that {\it a necessary condition for realizing 
the AdS/CFT correspondence 
is the presence of a 6D black string with traveling waves $p(u)$.} 
On the D-brane side, then we have the following picture. One finds 
$4Q_1Q_5$ bosonic fields and an equal number of fermionic fields on 
the circle(S$^1$). These degrees of freedom ($6Q_1 Q_5$ bosonic fields) are not 
changed as the coupling constant moves from the weak to the strong. 
One increases the string coupling $g_s$ to arrive at the near-horizon 
region and finally to form an event horizon at $r=0$. 
But all of string fields should have the same 
momentum distribution as in the black string: 
$p^{3/2} \gg r_0^2 \vert \dot p \vert$. 

It is reasonable from the previous discussion to consider 
that the CFT of D-string with a central charge $\tilde c = 6 Q_1 Q_5$  
is well-defined on the remote boundary. 
Also we can calculate the microscopic entropy at strong coupling.
Considering (\ref{cardy}) together with $\tilde c= 6Q_1 Q_5$ and 
the large $p(u)/\kappa^2$ limit ($N_a \gg Q_1 Q_5$), one recovers 
S$_{\rm CFT} = \sqrt{2 \pi Q_1 Q_5 } \int_0^L \sqrt{p/\kappa^2} du$. 
As a result of S$^{\rm 10D}_{\rm BH}=$S$_{\rm CFT}$, the AdS/CFT correspondence 
for the entropy is established on 
AdS$_3\times $S$^3\times$T$^4$/CFT$_2$. 

\section{Discussion}
We have shown that a realization of the AdS/CFT 
correspondence for the entropy is possible when the BTZ black hole is 
embedded in the string theory(explicitly, a 5D black hole). 
To recover S$^{\rm BTZ}_{\rm BH} = 2 \pi \rho_0 / 4 G_3$, we need 
both the central charge $c=3 \ell/ 2 G_3$ and the level number 
$N_K = \rho_0^2 / 4 G_3 \ell$. However, the pure 3D gravity provides 
us with $c=1$ (not $ c=3 \ell/ 2 G_3$). This is too small to obtain 
the Bekenstein-Hawking entropy. 
This is so because the 3D gravity is a pure gauge theory and examines 
only its macroscopic properties. 
Thus the correspondence between an exact AdS$_3$ and CFT$_2$ is not 
confirmed until now. 
On the other hand, when the BTZ black hole is embedded 
in a 5D black hole, one finds $\tilde c= 3 \tilde \ell / 2 \tilde G_3$ 
$=6 Q_1 Q_5$. This is exactly the sum of $4 Q_1 Q_5$-bosonic string fields 
and $4 Q_1 Q_5$-fermionic string fields on the D-brane side.  The total 
momentum number $N_K$, in the extremal case, is given by 
$L_0 = J = N_K$ and $\bar L_0 =0$, where $L_0$ and $\bar L_0$ are 
the generators of Virasoro algebra. However, this algebra is a realization 
of the asymptotic symmetry group of an exact AdS$_3$ at infinity($\rho=\infty$)
\cite{Bro86CMP207}. 
In general, the AdS/CFT correspondence is well-defined on the 
remote boundary at $\tilde \rho=\sqrt{\tilde \rho_0^2 + R^2}$. 
Hence we need the other mechanism 
which can transfer the information of the horizon 
($\tilde \rho = \tilde \rho_0$ : inner boundary) to the 
outer boundary ($\tilde \rho = \sqrt{\tilde \rho_0^2 + R^2}$).  
As far as we know, this exists only when the 
background geometry which interpolates between AdS$_3$ and flat space 
has a null Killing symmetry. The extremal 6D black 
string solution has such a symmetry and thus one adds  traveling waves 
with $p(u)$ to this solution. Then they transfer physical information 
(momentum distribution) from the horizon to the remote boundary. 
This is evident from the propagation of the graviton $p(u)$ in the 
extremal black string background. Also this comes from 
the fact that the non-AdS of (\ref{metric-10D}) with $p(u)$ can be 
transformed to AdS without $p(u)$. 
This implies that the near-horizon geometry is completely 
independent of the wave profile $p(u)$. Hence we suggest that 
{\it a necessary condition for realizing the AdS/CFT correspondence
is the presence of a 6D black string with traveling waves $p(u)$.}
Here $p(u)$ seems to be a messenger which communicates 
the information of the bulk(black string) to the boundary(CFT).
It plays an important role in counting the entropy, but not 
in the calculation of the greybody factor.

As a result, we provide 
a concrete model which shows a realization of the 
AdS/CFT correspondence for the entropy in the near-horizon. 
The other realization of AdS$_3$/CFT$_2$ correspondence 
was studied by using a 3D black string in 
Ref.\cite{Kal9804062}.

\section*{Acknowledgement}
This work was supported in part by the Basic Science Research Institute
Program, Minstry of Education, Project NO.
BSRI-98-2413 and grant from Inje University, 1998.

\end{document}